\documentclass[aps,prl,superscriptaddress,twocolumn]{revtex4-2}

\usepackage{graphics}
\usepackage{graphicx} 
\usepackage{dcolumn}
\usepackage{bm}
\usepackage{epstopdf}
\usepackage[usenames, dvipsnames]{color}
\usepackage{amsmath}

\begin{document}


\title{Local site behavior of the 5$d$ and 4$f$ ions in the frustrated pyrochlore Ho$_2$Os$_2$O$_7$}



\author{S.~Calder}
\email{caldersa@ornl.gov}
\affiliation{Neutron Scattering Division, Oak Ridge National Laboratory, Oak Ridge, Tennessee 37831, USA.}

\author{Z.~Y.~Zhao}
\affiliation{Materials Science and Technology Division, Oak Ridge National Laboratory, Oak Ridge, TN 37831.}

\author{M.~H.~Upton}
\affiliation{Advanced Photon Source, Argonne National Laboratory, Argonne, Illinois 60439, USA.}

\author{J.-Q.~Yan}
\affiliation{Materials Science and Technology Division, Oak Ridge National Laboratory, Oak Ridge, TN 37831.}


\begin{abstract}	
The pyrochlore osmate Ho$_2$Os$_2$O$_7$ is a candidate material for a fragile $J$=0 local singlet ground state, however little is known regarding the single-ion behavior of either the Os or Ho ions. To address this we present polarized neutron powder diffraction (PNPD) and resonant inelastic x-ray scattering (RIXS) measurements that separately probe the local site behavior of the Os and Ho ions. The PNPD results are dominated by Ho$^{3+}$ scattering and the analysis reveals local site susceptibility behavior consistent with spin ice materials. Complimentary unpolarized neutron powder diffraction show an ordered spin ice ground state in an applied magnetic field.  To isolate the Os$^{4+}$ single-ion behavior we present resonant inelastic x-ray scattering (RIXS) measurements at the osmium $L$-edge. Analysis of the RIXS spectra parameterize the spin-orbit coupling (0.35 eV), Hund's coupling (0.27 eV) and trigonal distortion (-0.17 eV). The results are considered within the context of a $J$=0 model and possible departures from this through structural distortions, excitonic interactions and 5$d$-4$f$ interactions between the Os ion and the surrounding Ho lattice. The experimental methodology employed highlights the complimentary information available in rare earth based 5$d$ pyrochlores from distinct neutron and x-ray scattering techniques that allow for the isolation and determination of the behavior of the different ions.  
\end{abstract}

\maketitle

\section{Introduction}

\begin{figure}[tb]
	\centering         
	\includegraphics[trim=0.5cm 4.5cm 0.5cm 0cm,clip=true, width=1.0\columnwidth]{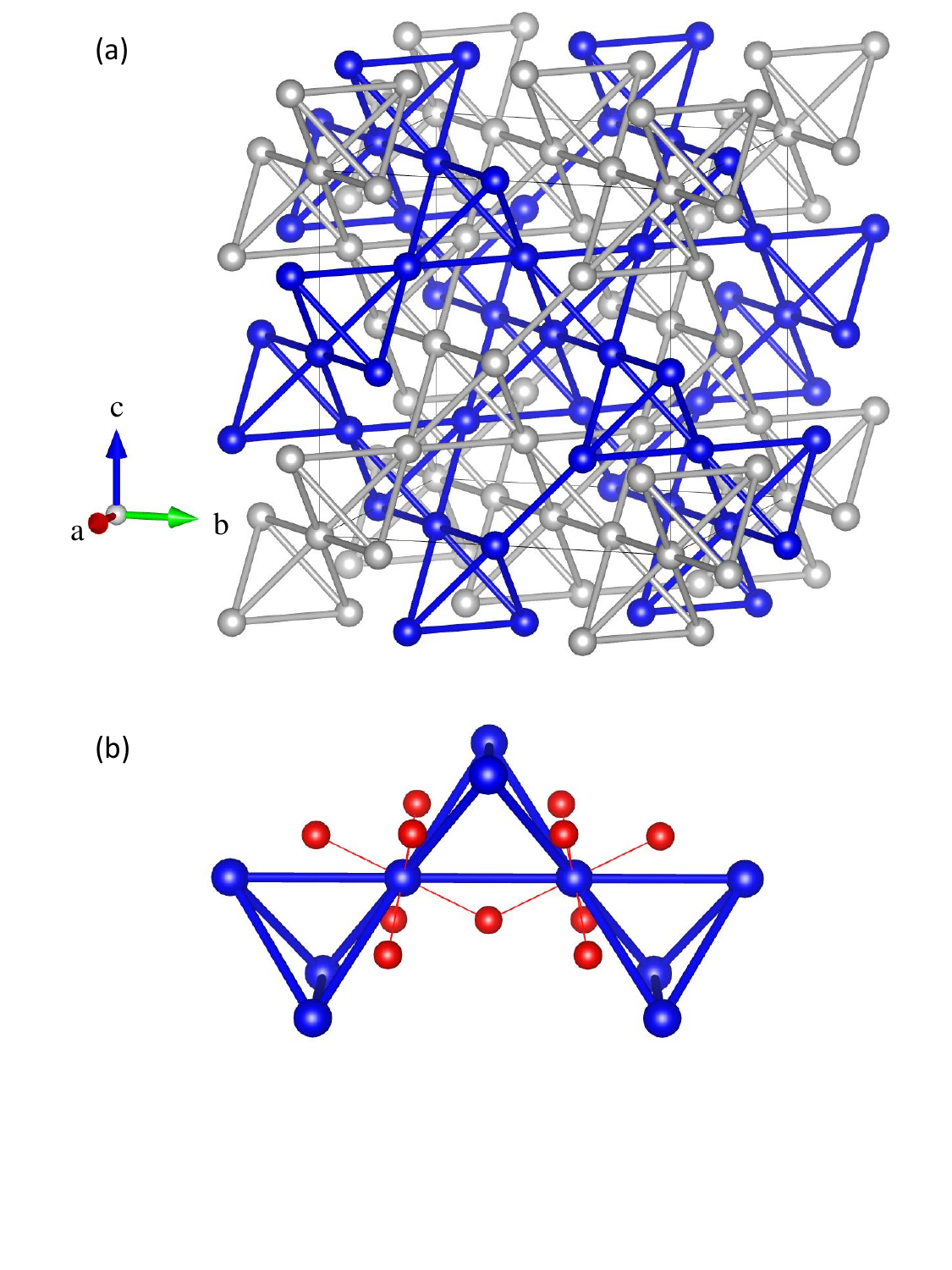}           
	\caption{\label{Fig_crystal_structure} (a) Interpenetrating lattices of corner sharing tetrahedra of Os (blue) and Ho (white) ions. The unit cell for the $Fd\bar{3}m$ space group is shown with the solid black lines. (b) Expanded view of Os and O (red) ions. The local symmetry of the Os ion results in a trigonally distorted crystal field environment.}
\end{figure}

Pyrochlore oxides with general formula  $A_2$$^{3+}B_2$$^{4+}$O$_7$ have formed an enduring interest in condensed matter research due to the lattice topology of a highly frustrated set of interpenetrating corner sharing tetrahedra on both the $A$ and $B$ sites (Fig.~\ref{Fig_crystal_structure}). A wide variety of physical phenomena have been predicted and observed in pyrochlores, including classic and quantum spin ice with emergent magnetic monopoles, spin liquid candidates, order by disorder, multipolar ordering, the anomalous Hall effect and metal-insulator transitions \cite{doi:10.1146/annurev-conmatphys-022317-110520, doi:10.1146/annurev-conmatphys-031016-025218, GREEDAN2006444, RevModPhys.82.53}. A family of pyrochlore compounds that has received only limited attention, but has the potential to host concomitantly intriguing behavior from the $A$ and $B$ sites is the rare earth osmates $R_2$Os$_2$O$_7$ ($R$=rare earth ion) \cite{SHAPLYGIN1973761}. The $A$-site $R$$^{3+}$ rare earth ion offers routes to investigate geometric magnetic frustration, and the $B$-site Os$^{4+}$ ion (5$d^4$) provides a candidate ion for singlet ground state magnetism \cite{PhysRevB.93.134426, PhysRevB.99.174442}. 

Investigations of singlet magnetism in materials with a $d^4$ ion originated in the 1960's \cite{JR9610003132} and has shown renewed interest in the context of strongly spin-orbit coupled (SOC) 4$d$/5$d$ magnetism \cite{doi:10.1146/annurev-conmatphys-020911-125138, doi:10.7566/JPSJ.90.062001}. When these $d$-ions are placed in an octahedral environment the 10$Dq$ crystal field splitting results in well separated $e_g$ and $t_{\rm 2g}$ levels. In the presence of sufficiently strong SOC there is a further splitting of the $t_{2g}$ levels into a $J_{\rm eff}=3/2$ quadruplet and a  $J_{\rm eff}=1/2$ doublet. The strong SOC in 5$d$ ions has been shown to provide candidates for this $J_{\rm eff}$ picture with $5d^5$-based iridates being a particular focus for $J_{\rm eff}=1/2$ quantum magnetism \cite{KimScience}. For $5d^4$ ions in these limits a $J_{\rm eff}=0$ ground state is predicted (henceforth referred to as $J$=0). Alternatively, consideration of strong Hund's coupling (J$_H$) leads to a total spin $S=1$ and effective $L_{\rm eff}=1$ orbital moment. These spin and orbital moments also result in a $J$=0 singlet when SOC is sufficiently strong. 

Novel magnetism emerging from the nominally non-magnetic $J$=0 state has been proposed from singlet-triplet excitations, where the ions develop collective magnetism due to interaction effects. In this model, excitations from the $J$=0 ground state into the J=1 state disperse within the crystal and condense into magnetically ordered states, if the exchange interactions overcome the energy gap between the $J$=0 and J=1 states \cite{PhysRevLett.111.197201, PhysRevB.91.054412}. Behavior consistent with a $J$=0 ground state has been observed in a small but growing number of materials including Sr$_3$Ir$_2$O$_7$F$_2$ \cite{PhysRevB.106.115140}, K$_2$RuCl$_6$ \cite{PhysRevLett.127.227201}, Ag$_3$LiRu$_2$O$_6$ \cite{PhysRevResearch.4.043079}, Sr$_3$
$M$IrO$_6$ ($M$=Na, Ag) and \cite{PhysRevMaterials.6.094415},  $A_2B$IrO$_6$ ($A$= Ba, Sr; $B$= Lu, Sc) \cite{PhysRevMaterials.6.094409},  and $M$$_2$YIrO$_6$ ($M$=Ba, Sr) \cite{PhysRevLett.112.056402, PhysRevB.96.064436}. The later cases of Sr$_2$YIrO$_6$ and Ba$_2$YIrO$_6$ double perovskites, however, have competing reports of magnetic and non-magnetic states questioning the applicability of the $J$=0 model \cite{PhysRevLett.120.237204, PhysRevB.101.014449}. This reinforces the need for a detailed characterization of candidate $J$=0 materials, particularity those with magnetism.

The rare earth pyrochlore osmates offer a potential route to investigate the $J$=0 scenario and can incorporate large 4$f$ magnetism on the $A$-site  \cite{SHAPLYGIN1973761}. The 5$d$ octahedra are corner sharing, rather than the more separated double perovskite octahedra. The superexchange interaction is therefore expected to be increased and provide a more amenable path for exchange driven exitonic magnetism predicted in $J$=0 materials \cite{PhysRevLett.111.197201, PhysRevB.91.054412}. Studies have been extremely limited on $5d^4$ pyrochlores, with reports on Y$_2$Os$_2$O$_7$ \cite{PhysRevB.93.134426, PhysRevB.99.174442} and Ho$_2$Os$_2$O$_7$ \cite{PhysRevB.93.134426}. While Y$_2$Os$_2$O$_7$ appears to not show any correlated magnetic order, there are indications that Ho$_2$Os$_2$O$_7$ may undergo a magnetic transition at 36 K \cite{PhysRevB.93.134426}. 

An important factor to consider in Ho$_2$Os$_2$O$_7$ is the magnetic Ho$^{3+}$ ion with potential for long or short ranged magnetism of the large moment. The related Ho$_2$Ti$_2$O$_7$ is a spin-ice material, where each local tetrahedra has two-spins pointing to the center and two spins pointing away from the center of the tetrahedra \cite{PhysRevLett.79.2554}. This ``two-in, two-out" spin ordering is a result of  geometric magnetic frustration of the Ho$^{3+}$ ions and leads to a highly degenerate ground state with fractionalized excitations of magnetic monopoles \cite{WOS:000252079300028}. Further realizations are Ho$_2$Sn$_2$O$_7$ and Ho$_2$Ge$_2$O$_7$, as well as the dysprosium analogues \cite{PhysRevLett.108.207206}. Strong local anisotropy of the Ho-ion leads to the conditions for the ``two-in, two-out" behavior, with the Ising spins being constrained to a local [111] axis. We show here that Ho$_2$Os$_2$O$_7$ provides a material in which to investigate spin-ice behavior within a candidate fragile $J$=0 state.

Understanding the single-ion behavior of both the Os and Ho ions in Ho$_2$Os$_2$O$_7$ can provide unique insights into materials with strong SOC, strong Hund's coupling, lattice distortions and spin-ice physics and their concomitant interplay. Here, we undertake an experimental investigation using the complementarity of polarized neutron diffraction and resonant x-ray scattering to separately access the rare earth and transition metal ion behavior in powder samples. The polarized neutron powder diffraction (PNPD) measurements reveal the local site susceptibility tensor of the Ho ion. As we show, this indicates behavior analogous to classical spin ice. To isolate the Os ion behavior and access the electronic ground state we perform resonant inelastic x-ray scattering (RIXS) at the Os $L$-edge. We parameterize the results in terms of SOC, Hund's coupling and trigonal distortion for the Os ion and discuss the implications on the magnetism in Ho$_2$Os$_2$O$_7$.

\section{Experimental Details}\label{Experimental_details}

\subsection{Sample synthesis}\label{Samples}

The sample synthesis and characterization for Ho$_2$Os$_2$O$_7$ has been described previously \cite{PhysRevB.93.134426}. Ho$_2$Os$_2$O$_7$ contained Ho$_2$O$_3$ and OsO$_2$ impurity phases. These impurities did not affect the determination of microscopic magnetic properties in Ref.~\onlinecite{PhysRevB.93.134426} and the methodology of polarized neutron and resonant x-ray scattering are anticipated to be less influenced by these phases due to being able to select the main phase scattering.

\subsection{Neutron powder diffraction in a magnetic field}\label{Mag_field_HB2A}

Neutron powder diffraction measurements on 1.6 grams of Ho$_2$Os$_2$O$_7$ were carried out on the HB-2A powder diffractometer at the High Flux Isotope Reactor (HFIR), Oak Ridge National Laboratory (ORNL) \cite{Garlea2010, doi:10.1063/1.5033906}. These measurements were performed in a cryomagnet in fields up to 3 T in the temperature range 1.5 K to 100 K. The sample was pressed into pellets of 6 mm diameter to ensure no movement of the powder grains when the magnetic field was applied. This maintains the powder average and allows for analysis of the data. The pellets were contained in an Al sample holder. Constant wavelength measurements were performed at 2.41 $\rm \AA$ from the Ge(113) monochromator reflection. The pre-mono, pre-sample and pre-detector collimation was open-open-12'. A pyrolytic graphite (PG) filter was placed before the sample to remove higher order reflections for the 2.41 $\rm \AA$ wavelength. The diffraction pattern was collected by scanning a 120$^{\circ}$ bank of 44 $^3$He detectors in 0.05$^{\circ}$ steps to give 2$\theta$ coverage from 5$^{\circ}$ to 130$^{\circ}$. Rietveld refinements were performed with Fullprof \cite{Fullprof}. Symmetry allowed magnetic structures were considered using both representational analysis with SARAh \cite{sarahwills} and magnetic space groups with the Bilbao Crystallographic Server \cite{Bilbao_Mag}. Plots of the crystal and magnetic structure were prepared using VESTA \cite{VESTA}.

\subsection{Polarized Neutron Powder Diffraction (PNPD)}\label{PNPD}

Polarized neutron powder diffraction (PNPD) was also performed on the HB-2A powder diffractometer with a constant wavelength of 2.41 $\rm \AA$. A supermirror V-cavity was placed between the monochromator and the sample to select and transmit one neutron polarization state. A guide field ``flipper" after the V-cavity allowed the polarization state of the neutron beam on the powder sample to be controlled between spin-up and spin-down states. The sample, in pressed pellet form to avoid grain reorientation, was placed in an Al sample holder and contained within a vertical field asymmetric cryomagnet. A detector bank consisting of 44 $^3$He point detectors over a range of 120 degrees collected the scattered neutron intensity. No polarization analysis was performed after the sample. Measurements were taken with the spin up and spin down incident beam for temperatures of 5 K, 15 K, 25 K, 30 K, 40 K and 50 K in an applied field of 1 T. A measurement was taken at 0 T and 50 K to provide a zero field unpolarized reference for the data analysis. The data analysis was performed using the open source CrysPy software  to determine the local site susceptibility tensor \cite{CrysPy}.

\subsection{Resonant inelastic X-ray scattering (RIXS)}\label{RIXS}

Powder samples of Ho$_2$Os$_2$O$_7$ were measured with RIXS at the Os $L_3$-edge (10.88 keV) on Sector-27 at the Advanced Photon Source (APS) using the MERIX instrumentation \cite{SHVYDKO2013140}. The incident energy was accessed with two pre-sample monochromators, a primary Diamond(111) monochromator and a secondary Si(400) monochromator. The energy of the beam scattered from the sample was discriminated with a Si(446) 2m diced analyzer. The detector was a MYTHEN strip detector. To reduce the elastic line we performed inelastic measurements in horizontal geometry at 2$\theta$=90$^{\circ}$. The RIXS energy resolution was 130meV FWHM. The powder sample was held in an aluminum mount with a trench for the powder and Kapton paper used to cover and seal the powder. Measurements were taken at room temperature and base temperature of the closed-cycle refrigerator of 6 K. The data analysis was performed using the open source EDRIXS software that allows for calculations of RIXS spectra of strongly correlated materials based on exact diagonalization (ED) of model Hamiltonians  \cite{WANG2019151}. This was used to determine to determine the crystal field ground state of the Os$^{4+}$ ion.

\begin{figure}[tb]
	\centering         
	\includegraphics[trim=0.5cm 3cm 3cm 0cm,clip=true, width=0.86\columnwidth]{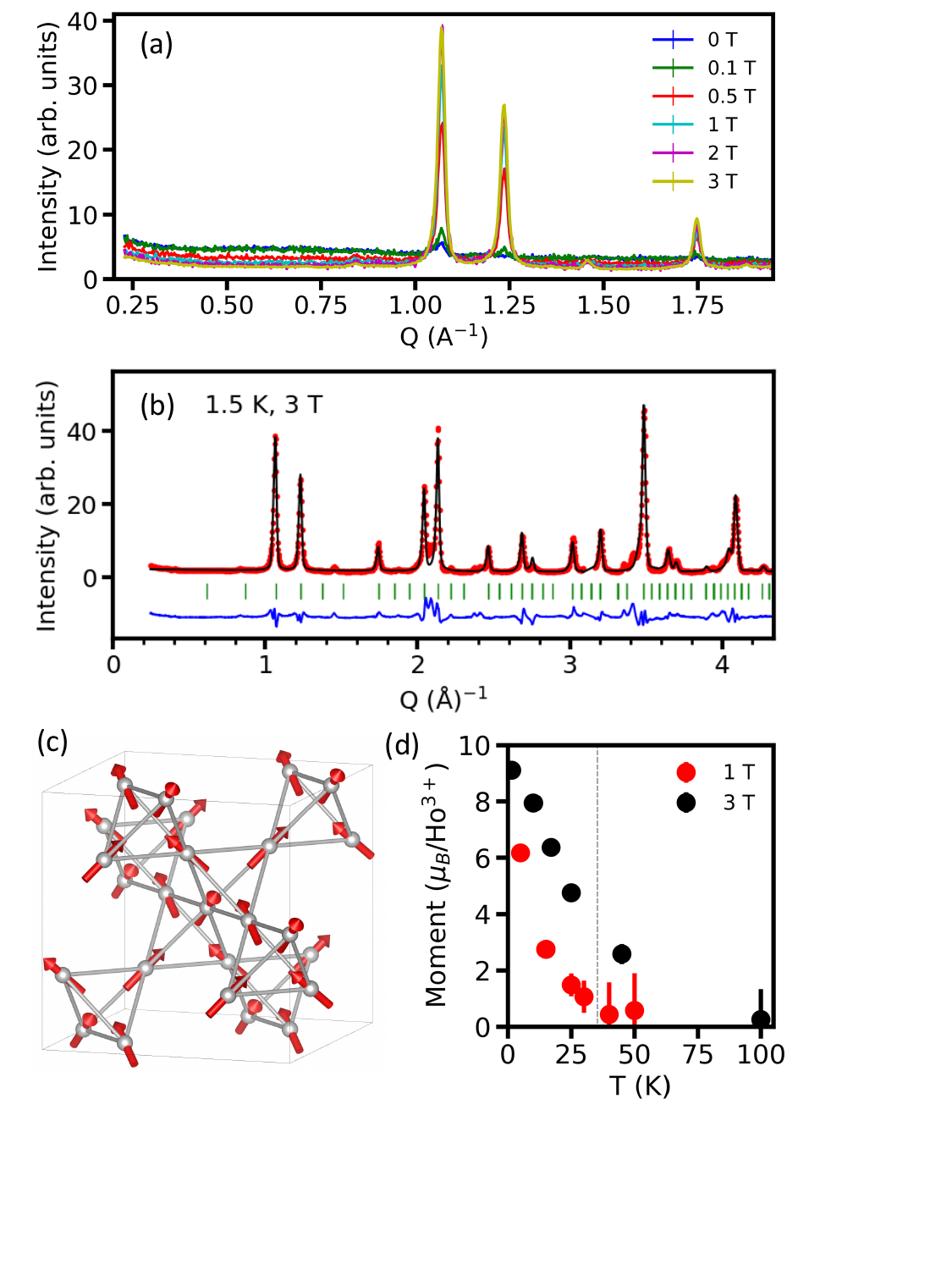}           
	\caption{\label{Fig_HB2A_MagField}(a) Neutron powder diffraction data at 1.5 K in applied magnetic fields. (b) Refinement to a ordered spin ice model at 1.5 K and 3 T. Red is data, black is model, blue is the difference and the green ticks are the reflections. (c) The ordered spin ice structure shown for the Ho ions within a single magnetic unit cell. (d) Refined moment sizes of the Ho ion at various temperatures for 1 T and 3 T fields. The dashed grey vertical line is at 36 K to highlight the onset of correlations observed at 0 T.}
\end{figure}

\section{Results and Discussion}

\subsection{Applied Magnetic Field Neutron Powder Diffraction}\label{MagField_HB2A}

We begin with results from neutron powder diffraction measurements on Ho$_2$Os$_2$O$_7$. Previous zero-field neutron scattering measurements revealed diffuse scattering developing below 36 K  \cite{PhysRevB.93.134426}. This indicates short range correlations that could naturally arise from the frustrated lattice geometry, as observed in several rare earth pyrochlores \cite{GREEDAN2006444, RevModPhys.82.53}. The large magnitude of the signal is only compatible with this being from the Ho ion, although potentially induced by ordering on the Os site. 

To gain further insights an applied magnetic field measurement was performed at several fields and temperatures. Neutron powder diffraction data collected at 1.5 K in fields from 0 T to 3 T are shown in Fig.~\ref{Fig_HB2A_MagField}(a). The effect of the magnetic field is to change the diffuse scattering from short-range order at 0 T to long-range ordering. These magnetic reflections can all be indexed to nuclear reflection positions, indicating a {\bf k}=(000) propagation vector. The high temperature crystallographic space group for Ho$_2$Os$_2$O$_7$, $Fd\bar{3}m$, with {\bf k}=(000)  has 4 maximal magnetic space groups \cite{Bilbao_Mag}. These are, in BNS notation, $Fd\bar{3}m'$ ($\# 227.131$), $I4_1/am'd'$ ($\#$141.557), $I4'_1/amd'$ ($\#$141.555)  and $Imm'a'$ ($\#$74.559). These allow variously "all-in, all-out" order, ordered spin-ice, order in a fixed $ab$-plane, as well as more complex order. Refining the data using the different magnetic space groups showed that only the magnetic space group $I4_1/am'd'$ ($\#$141.557) was able to match the data. This corresponds to an ordered spin-ice magnetic structure for the Ho ions. The refinement is shown in Fig.~\ref{Fig_HB2A_MagField}(b) and the corresponding spin structure is shown in Fig.~\ref{Fig_HB2A_MagField}(d).

This ordered spin-ice state is found in magnetic field studies of  Ho$_2$Os$_2$O$_7$,  Dy$_2$Os$_2$O$_7$ \cite{PhysRevB.72.224411} and  for zero-field in Tb$_2$Sn$_2$O$_7$  \cite{PhysRevLett.94.246402}. Pyrochlores with two magnetic ions are also observed to form this long-range ordered spin ice ground state. Sm$_2$Mo$_2$O$_7$ \cite{PhysRevB.77.020406} and Nd$_2$Ir$_2$O$_7$ \cite{PhysRevLett.115.056402} can be induced with an applied magnetic field, while Nd$_2$Mo$_2$O$_7$  forms this ground state in zero field \cite{doi:10.1143/JPSJ.70.284}.

The refinement of the magnetic structure at 1.5 K and 3 T is shown in Fig.~\ref{Fig_HB2A_MagField}(b). The ordered spin ice structure can be viewed in Fig.~\ref{Fig_HB2A_MagField}(c). The ordered moment size is 8.1(1) $\rm \mu_B$/$\rm Ho^{3+}$ ion. This is close to the fully saturated ordered moment of 10 $\rm \mu_B$/$\rm Ho^{3+}$ for an unperturbed J = 8 Ho$^{3+}$ ion.

The temperature evolution of the magnetic moment in Fig.~\ref{Fig_HB2A_MagField}(d) shows the development of long range order in a field of 1 T occurs around the same temperature as the onset of the diffuse scattering of 36 K for zero-field. This onset is shifted to higher temperatures for the higher magnetic field measurements of 3 T.

\begin{figure}[tb]
	\centering         
	\includegraphics[trim=0cm 0cm 10.0cm 0cm,clip=true, width=0.72\columnwidth]{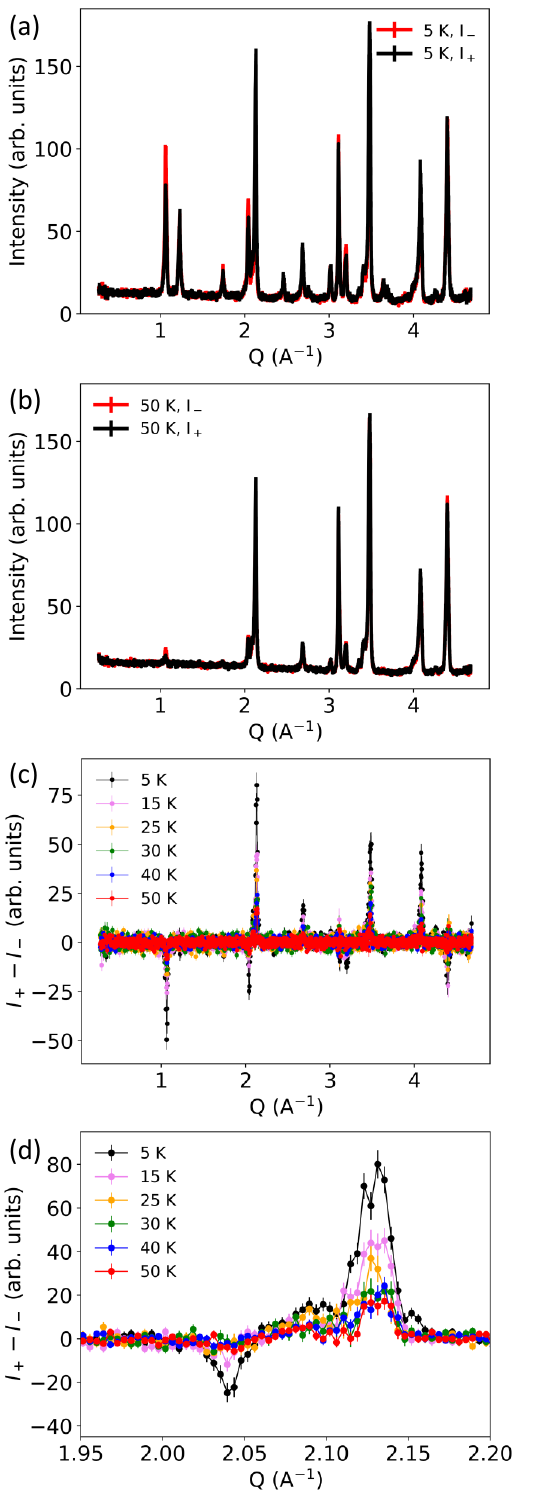}           
	\caption{\label{Fig_polarized_data} Polarized neutron powder diffraction data measured with spin up (I$_+$) and spin down (I$_-$) incident polarization at (a) 5K, 1 T and (b) 50 K, 1 T.  (c) The difference of spin up minus spin down ($I_+$-$I_-$) measurements  for temperatures 5 K, 15 K, 25 K, 30 K, 40 K and 50 K. (d) Plot over a select Q range to allow for clearer visualization of the temperature dependence of the difference.}
\end{figure} 

\begin{figure}[tb]
	\centering         
	\includegraphics[trim=0cm 0cm 10cm 0cm,clip=true, width=0.72\columnwidth]{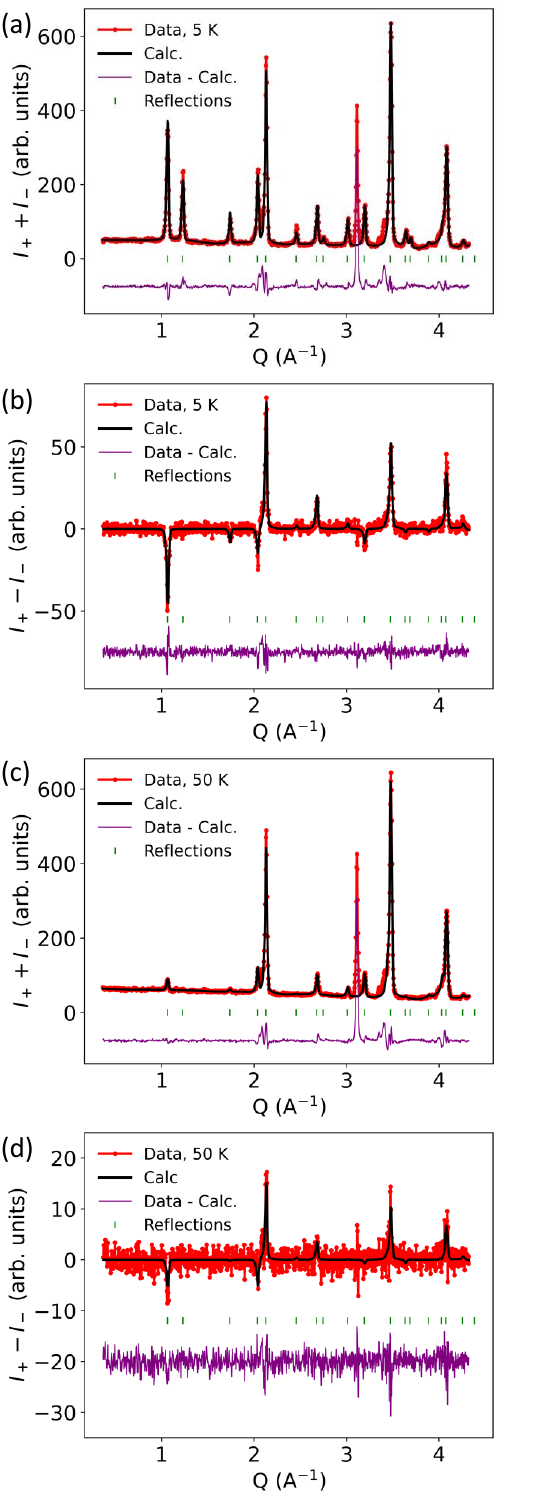}           
	\caption{\label{Fig_polarized_Fits} Refinement of the polarized neutron powder diffraction data collected on HB-2A for the (a) sum ($I_+$+$I_-$) at 5 K, (b) difference ($I_+$-$I_-$) at 5 K, (c) sum ($I_+$+$I_-$) at 50 K and (d) difference ($I_+$-$I_-$) at 50 K. The peak at 3.1 $\rm \AA^{-1}$ in the $I_+$+$I_-$ data was from the Al sample holder and was excluded during the fitting.}
\end{figure}

\subsection{Local site susceptibility tensor with Polarized Neutron Powder Diffraction (PNPD)}\label{Polarized}

To directly access the Ho single ion behavior we extended the neutron investigation with polarized neutron powder diffraction (PNPD) measurements on Ho$_2$Os$_2$O$_7$. As with the unpolarized measurements on Ho$_2$Os$_2$O$_7$, it is anticipated that the large Ho$^{3+}$ moment will overwhelm any potential signal from the Os$^{4+}$ ion. 

PNPD provides a direct measure of the local site susceptibility tensor to reveal the local anistropy of the ion, which can be central to the emergent behavior. The local site susceptibility tensor can be determined from measurements of a sample in an applied magnetic field with  the incident neutron polarization state either parallel ($I_+$) or anti-parallel ($I_-$) to the field direction. This technique is described in detail in Refs.~\onlinecite{Gukasov_2010, PhysRevResearch.1.033100}. For rare earth pyrochlores the information obtained from powders is often equivalent to that for single crystals \cite{PhysRevLett.103.056402, PhysRevResearch.1.033100}, making PNPD a powerful technique to obtain local anisotropic behavior.

The flipping difference method is used here, with the intensity difference of spin up ($I_+$)  and spin down ($I_-$) PNPD measurements given by:

\begin{eqnarray}
\Delta I = I_+ - I_- \propto 2 \Re [ F_N^\star \langle {\bf F}_{M,\perp}  \cdot  {\bf P} \rangle   ]  ,
\end{eqnarray}

where $F_N$ and ${\bf F}_M$ are the nuclear and magnetic structure factors and $\bf{P}$ is the polarization value of the incident beam. Angular brackets account for the powder averaging. This relationship reveals that the difference signal is only observed in PNPD when there is a magnetic signal and also that the signal is, counter-intuitively, enhanced at {\it larger} nuclear reflections. The magnetic structure factor is given by ${\bf F}_{M}(Q)=\sum_i {\bf m_i}f_m(Q) \exp(iQ.r_i)$, where the sum is over the unit cell, $f_m(Q)$ is the magnetic form factor and ${\bf m_i}$ is the magnetic moment on atom $i$. ${\bf m_i}$ in a magnetic field satisfies the behavior:

\begin{eqnarray}
\bar{\chi_i} = \frac{{\bf m_i}}{{\bf B}} = \begin{pmatrix}
{\chi_{11}} & {\chi_{12}} & {\chi_{13}}\\
{\chi_{12}} & {\chi_{22}} & {\chi_{23}}\\
{\chi_{13}} & {\chi_{23}} & {\chi_{33}}\\
\end{pmatrix}	,
\end{eqnarray}

The site symmetry determines any constraints and the allowed non-zero tensor components $\chi_{ij}$.  

\begin{figure}[tb]
	\centering         
	\includegraphics[trim=0cm 11cm 0cm 0cm,clip=true, width=1.0\columnwidth]{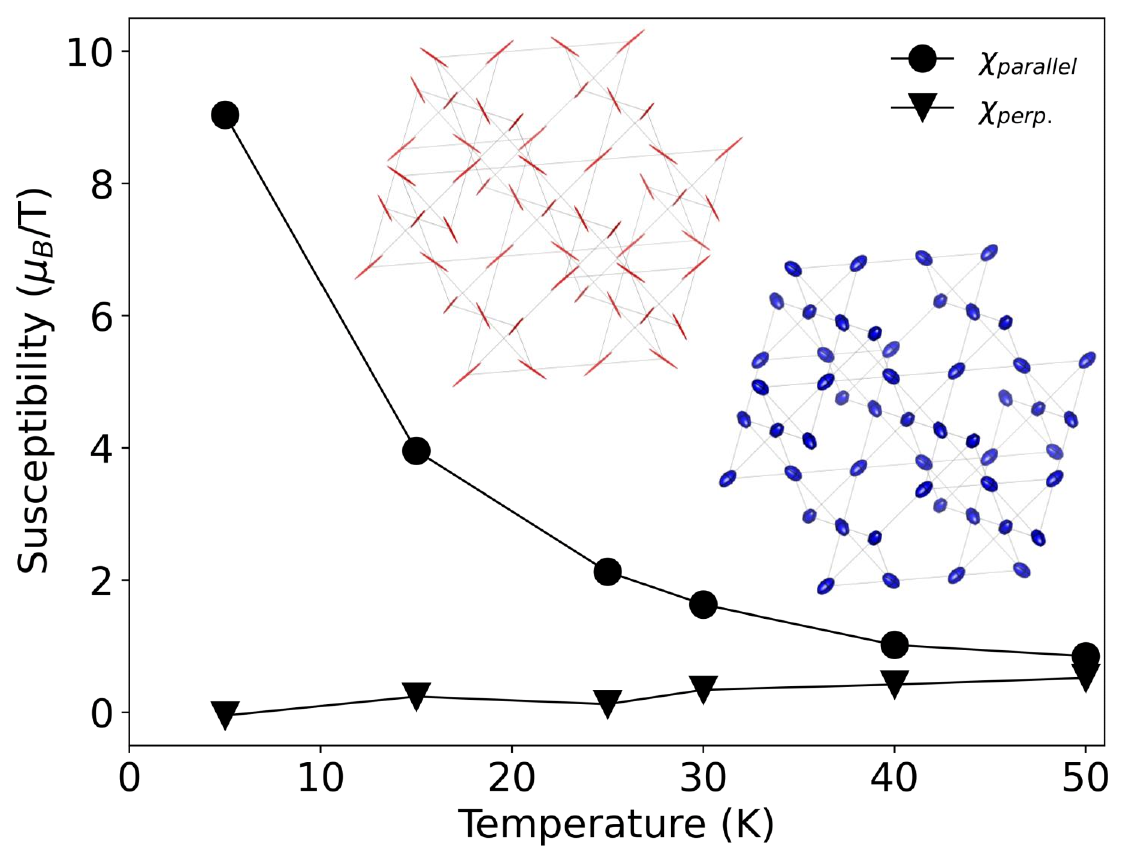}           
	\caption{\label{Fig_PNPD_Susceptibility} Temperature dependence of the measured susceptibility ($\chi$) of Ho$_2$Os$_2$O$_7$ for $\chi_{parallel}$ and $\chi_{perp.}$. The inset shows the Ho magnetization ellipsoids at 5 K (red) and 50 K (blue).}
\end{figure} 

$\bar{\chi_i}$ can be extracted directly from PNPD measurements. The data collected with spin up ($I_+$) and spin down ($I_-$) polarized incident neutron beam at 5K, 1T and 50K, 1 T are shown in Figs.~\ref{Fig_polarized_data}(a)-(b). The intensity changes with polarized state and this can be viewed for all temperatures by considering the flipping difference ($\Delta I$=$I_+ - I_-$) for all temperatures measured in Figs.~\ref{Fig_polarized_data}(c)-(d). The difference intensity shows a strong temperature response.

To extract quantitative information the polarized data was refined using the CrysPy software to extract $\bar{\chi_i}$ \cite{PhysRevResearch.1.033100}. This is shown in Figs.~\ref{Fig_polarized_Fits}(a)-(d). Inspection of the data reveals that certain reflections show a clear difference in intensity for different polarized states, while other reflections show no observable change. For example, consider the two peaks at 1.06 $\rm \AA^{-1}$ and 1.23 $\rm \AA^{-1}$, corresponding to the (111) and (200) reflections, respectively. The (111) reflection shows a large intensity change in the difference and sum, whereas for the (200) reflection there is a change in the sum, but no signal observed in the difference. This was utilized in the data analysis by refining both the  sum ($I_+ + I_-$) and the difference ($I_+ - I_-$).

For the case of the Ho ion at the 16$d$ Wyckoff position in $Fd\bar{3}m$ space group the symmetry constraints lead to $\chi_{11}=\chi_{22}=\chi_{33}$ and $\chi_{12}=\chi_{13}=\chi_{23}$. Consequently there are only two independent variables in the matrix tensor with the principle axes of the magnetization ellipsoids along the four local $\left\langle 111 \right\rangle$ directions. We follow the treatment in Refs. \onlinecite{PhysRevLett.103.056402, PhysRevResearch.1.033100} to define $\chi_{parallel}$ and $\chi_{perp.}$ terms as $\chi_{parallel}$=$\chi_{11}+2\chi_{12}$ and $\chi_{perp.}$=$\chi_{11}-\chi_{12}$ and plot these values for each temperature in Fig.~\ref{Fig_PNPD_Susceptibility}. There is a strong temperature dependence of $\chi_{parallel}$/$\chi_{perp.}$, which indicates a change in the local anisotropy. The local site susceptibility can be visualized as magnetization ellipsoids around the ion, which is shown as the inset in Fig.~\ref{Fig_PNPD_Susceptibility} for 5 K (red ellipsoids) and 50 K (blue ellipsoids). At the lowest temperature measured of 5 K this density shows highly anisotropic Ising behavior, with the magnetization density contrained to the local $\left\langle 111 \right\rangle$ directions. As the temperature increases the behavior becomes isotropic. The $\chi_{parallel}$ and $\chi_{perp.}$ behavior was reported for several rare earth titanates from polarized measurements \cite{PhysRevLett.103.056402}. The results in Fig.~\ref{Fig_PNPD_Susceptibility} for Ho$_2$Os$_2$O$_7$ show the expected response for canonical spin ice (Ho$_2$Ti$_2$O$_7$) and is distinct from other frustrated pyrochlore magnetism \cite{PhysRevResearch.1.033100, PhysRevLett.103.056402}.

The neutron scattering results allow Ho$_2$Os$_2$O$_7$ to be placed into the category of a new candidate spin-ice material. An analogy can be made with the related $4d^4$ based pyrochlore Ho$_2$Ru$_2$O$_7$, which was also considered as a candidate spin ice. Measurements showed, however, that the  Ru$^{4+}$ and  Ho$^{3+}$ ions order at 96 K and 1.4 K, respectively, in zero field \cite{PhysRevLett.93.076403}. At the ordering temperature of the Ru$^{4+}$ ions there is concomitant short range order of the Ho ions, before the eventual long range order of the Ho sublattice at 1.4 K. Subtle crystal field effects at the 96 K transition and complex superexchange interactions between the 4$f$ and 4$d$ ions drive this behavior in Ho$_2$Ru$_2$O$_7$ \cite{PhysRevLett.93.076403}. For Ho$_2$Os$_2$O$_7$ no such breaking of the fluctuating spin ice state is observed from zero field neutron diffraction measurements, with diffuse scattering remaining present to 0.3 K \cite{PhysRevB.93.134426}. As we show this highly degenerate spin ice ground state can be broken with the application of a magnetic field into an order spin ice state. Further measurements at lower temperatures would be of interest to explore the potential for any zero field long range order and contrast this with the observed ordering in a magnetic field.

\subsection{5d$^4$ ground state of the Os ion in Ho$_2$Os$_2$O$_7$ measured with RIXS}\label{RIXS}

\begin{figure}[tb]
	\centering         
	\includegraphics[trim=0cm 0cm 8cm 0cm,clip=true, width=0.86\columnwidth]{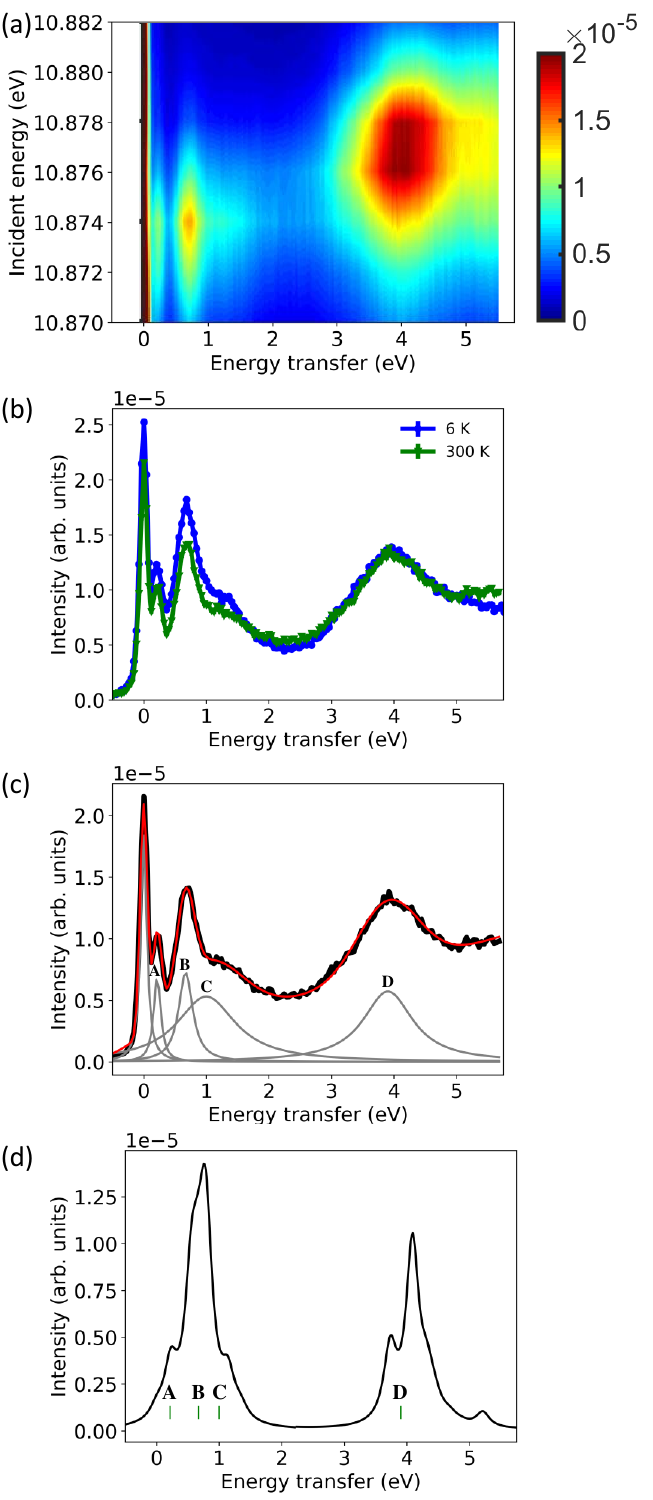}           
	\caption{\label{Fig_RIXS_Ho227}(a) RIXS map of energy transfer as the incident energy is varied through the Os $L_3$-edge at 300 K. (b) Energy transfer measured at 6 K and 300 K with incident energy $\rm E_i=10.874$ keV. (c) To extract the energy of the observed peaks in the energy transfer spectra the data was fit with Lorentzians. The data is the black circles, the overall fit is the red line and the individual Lorentzians used are shown as the grey lines. (d) Model of the energy loss spectra.}
\end{figure}

To investigate the Os single ion behavior we turn to RIXS measurements. RIXS has become a standard tool to extract details on the electronic ground state and, in the case of 5$d$ ions, is better suited to the energy scales over other techniques, such as inelastic neutron scattering which becomes unfeasible for 5$d$ ions due to signal and resolution constraints. The RIXS measurements presented were carried out at the Os $L_3$ edge, which enhances and isolates the Os signal. A map of the intensity of the inelastic energy loss as the incident energy ($\rm E_i$) was tuned through the Os $L_3$ edge in Ho$_2$Os$_2$O$_7$ is shown in Fig.~\ref{Fig_RIXS_Ho227}(a) for a temperature of 300 K. Two distinct inelastic regions are observed with different incident energy resonant behavior: below 2 eV and above 2 eV. The scattering below 2 eV has a resonant energy of 10.874 keV, whereas the scattering above 2 eV energy loss has a resonant energy of 10.8775 keV. The resonant energies correspond to the core-hole transition energy accessed during the RIXS process. For Os $L$-edge RIXS the core-hole process is 2$p$-5$d$. The 5$d$ manifold, however, is nominally split into $t_{2g}$ and $e_g$ sub-manifolds that results in a distinction between the energies of the 2$p$-5$d$($t_{2g}$) and 2$p$-5$d$($e_g$) resonant processes. The scattering involving excitations within the $t_{2g}$ manifold will occur at a lower energy than scattering involving $e_g$ levels, with the energy difference corresponding to the $t_{2g}$ and $e_g$ splitting. This allows for the assignment of features below 2 eV as being due to intra-$t_{2g}$ processes and above 2 eV as involving an $e_g$ process. In addition it reveals the 10$Dq$ crystal field splitting of $t_{2g}$ and $e_g$ of the order $\sim$3.5 eV.

To follow the temperature dependence, measurements at an incident energy of 10.874 keV were performed at 6 K and 300 K, see Fig.~\ref{Fig_RIXS_Ho227}(b). While there is a change in the scattering between 300 K and 6 K, no resolvable change of the inelastic peak positions in energy are observed. In addition no new scattering that could be assigned to magnetic excitations were observed at 6 K. The overall change in the RIXS spectra intensity may be an experimental artifact of the beam scattering off different powder grains or could be an intrinsic effect in the sample indicating subtle electronic changes of the Os ion related to magnetic ordering at 36 K. Single crystal studies would be of interest to probe this behavior, as well as search for lower energy magnetic features beyond the limits of these powder measurements that could provide insights into any exitonic magnetism.

The RIXS spectra in Fig.~\ref{Fig_RIXS_Ho227}(b) have three resolvable inelastic peaks below 2 eV, followed by broad scattering above 2 eV. In Fig.~\ref{Fig_RIXS_Ho227}(c) we fit each of these (labeled A,B,C,D), along with the elastic line, using five Lorentzian functions to extract the energy and width. The results are shown in Table \ref{RIXS_Elevels_Ho227}. We note that apart from peak A the width of the fitted peaks indicate that they likely consist of more than one unresolvable peak.

\begin{table}[tb]
	\caption{\label{RIXS_Elevels_Ho227}Energy levels and width ($\Gamma$) from fitting the RIXS data to Lorentzians of the form ($A\sigma^2)/(\Gamma^2 + (x - E)^2$).}
	\begin{tabular}{| c | c | c |}
		\hline 
		Peak & Expt. Energy (eV)  & Width of peak $\Gamma$ (eV) \\ \hline
 Elastic	&	0  &   0.06     \\
A	&	0.21  &  0.07   \\
B	&	0.67   &   0.14  \\
C	&	1.0   &   0.57 \\
D	&	3.9   &   0.48  \\
		\hline
	\end{tabular}
\end{table}

We utilize the EDRIXS software to extract the energy scales of the competing parameters of SOC, Hund's coupling and lattice distortions controlling the single ion ground state \cite{WANG2019151}. To simulate scattering for an Os$^{4+}$ ion in Ho$_2$Os$_2$O$_7$ we include SOC, Hund's coupling, and Coulomb interactions for the case of four $d$ electrons on the Os ion. Utilizing a cubic crystal field ground state and fixing 10$Dq$=3.9 eV, corresponding to the center of the broad scattering at D in Fig.~\ref{Fig_RIXS_Ho227}(c), well matched the RIXS data for the region above 2 eV. This confirmed that the scattering $>$ 2eV can be nominally assigned to $e_g$ scattering. Attempting to model the RIXS data below 2 eV with a cubic ground state could not reproduce the positions of the observed scattering at peaks A, B and C. To go beyond a cubic 10$Dq$ crystal field approximation we additionally model trigonal distortions. This is applicable for Ho$_2$Os$_2$O$_7$ based on the distortion away from cubic due to the O1 (48$f$) oxygen position in the  $Fd\bar{3}m$. For this case the $x$ position of the O1 ion can vary from the cubic value of  $x$=0.3125. For  Ho$_2$Os$_2$O$_7$ the O1 $x$ value is 0.336 at 4 K. An increase in the $x$ value from ideal cubic is due to a trigonal compression of the octahedra in Ho$_2$Os$_2$O$_7$. Adding in a trigonal compression term to further split the ground state multiplets provided suitable models for the RIXS data. 

A model is presented in Fig.~\ref{Fig_RIXS_Ho227}(d) with SOC=0.35eV, Hund's coupling $\rm J_H$=0.27 eV and trigonal distortion of $\rm ds=-0.18$ eV. This broadly matches the energy and intensity variation of the RIXS spectra. Direct fitting of the spectra was not feasible, partly due to discrepancies in the intensity from the powder measurements. Nevertheless a narrow region of applicable solutions were determined to be $0.3$$\leq$SOC(eV)$\leq$$0.37$, $0.2$$\leq$$\rm J_H(eV)$$\leq$$0.35$ and $-0.2$$\leq$ds(eV)$\leq$$-0.15$. Values outside these parameters did not produce RIXS scattering at the observed energy positions from the data.  

Considering the ground state there are 15 $t^4_{2g}$ states for a $d^4$ ion. Large crystal field, interorbital Coulomb interactions and Hund's coupling lift this degeneracy to create a ninefold degenerate $^3T_{1g}$ state, described by an effective orbital moment $L$ = 1 and $S$ = 1. For strong SOC the $^3T_{1g}$ state is further split into a J=0 ground state with 3-fold ($\Gamma_4$) J=$1$ and 5-fold ($\Gamma_3$,$\Gamma_5$) $J$=2 excited states. These can be assigned to the elastic peak ($J$=0), peak A ($J$=1) and peak B ($J$=2) within this model. The $J$=1 excited state occurs at $\sim$SOC/2 and the $J$=2 excited state occurs at $\sim$3*SOC/2. The $^1T_{2}$, $^1E$, which have an energy of $\sim$2$J_H$ corresponds to peak C. The $^1A_1$ excited state, with expected $\sim$5$J_H$ is not observed and could be obscured by the tail from the broad scattering above 2 eV. Peak D corresponds to $t_{2g}$-$e_g$ excitations, nominally from the $10Dq$ splitting.  We note that further mixing can occur from excitations to  the $e_g$ states, which may be significant given the relatively similar energy scale of this excitation centered around 3.9 eV.

These results on Ho$_2$Os$_2$O$_7$ show agreement with the related  Y$_2$Os$_2$O$_7$ in a separate RIXS study \cite{PhysRevB.99.174442}. For Y$_2$Os$_2$O$_7$ $J$=0 behavior was assigned, although different analysis methodologies were employed with no direct comparison to the RIXS spectra. In a related study on the halides K$_2$OsCl$_6$, K$_2$OsBr$_6$, and Rb$_2$OsBr$_6$ the RIXS spectra showed no splitting of the spectra from any lattice distortion \cite{PhysRevB.108.125120}. This is distinct from the Ho$_2$Os$_2$O$_7$, and  also Y$_2$Os$_2$O$_7$ \cite{PhysRevB.99.174442}, spectra which show peaks that are broader than resolution due to splitting. Therefore the halides can be placed closer to the $J$=0 limit than the rare earth pyrochlores, indicating a more fragile state in Ho$_2$Os$_2$O$_7$. Despite this there are observations of weak magnetism in these $5d^4$ halides \cite{doi:10.1021/acs.inorgchem.2c02171}.

The parameters determined from modeling the RIXS spectra for Ho$_2$Os$_2$O$_7$ are physically reasonable and indicate large SOC and $\rm J_H$ values that are central to creating the electronic ground state in a single ion model. The role of the small but observable trigonal distortions on any potential induced magnetism is an important open question motivated by the RIXS results. Both the trigonal distortion and exchange driven exitonic magnetism are available mechanisms to induce magnetism and create the Os order indicated from previous susceptibility measurements \cite{PhysRevB.93.134426}. Going beyond the single ion picture and considering interactions between the 5$d$ and 4$f$ ions is furthermore required for a more complete picture of the behavior in Ho$_2$Os$_2$O$_7$.

\section{Summary and Conclusions}

Through a combined investigation of Ho$_2$Os$_2$O$_7$ with polarized neutron powder diffraction and resonant inelastic x-ray scattering new advances in the understanding of the local site behavior have been presented. The Ho$^{3+}$ and Os$^{4+}$ ions in this material sit on interpenetrating frustrated lattice with corner sharing tetrahedra and both show distinct behavior, with potential for coupling and interactions between the 5$d$ and 4$f$ ions. 

In the case of Ho$^{3+}$ the PNPD at low temperature shows a highly anisotropic magnetization density, as required for spin ice behavior. A spin ice model is consistent with observations of diffuse scattering in previous powder neutron diffraction measurements at 36 K that indicate a fluctuating ground state of frustrated spins that do not undergo long range magnetic order down to at least 0.3 K. Neutron diffraction measurements in a magnetic field presented here show that this short-range ordering can be driven into an ordered spin ice state in small magnetic fields. The ordering occurs around 36 K for 1 T, consistent with the zero-field onset of short range ordering, and is pushed to higher temperatures as the field is increased. Further measurements on Ho$_2$Os$_2$O$_7$ will be of interest to probe for exotic behavior associated with this candidate spin ice physics.

Using RIXS measurements of the Os ion allowed for a quantitative determination of SOC (0.35 eV) and Hund's coupling (0.27 eV) values, which are large and play an important role in creating the single ion ground state. The trigonal distortion intrinsic in the pyrochlore structure had to be included to best capture the measured spectra, with the ideal cubic approximation proving insufficient. Broadly the spectra is consistent with a $J$=0 state and associated $J$=1 and $J$=2 excited states, however further measurements are of interest to probe the influence the surrounding Ho lattice plays on any magnetic behavior of the Os ion beyond this single ion picture. This is particularly pertinent for 5$d$ ions that can have strong hybridization of the 5d orbitals with surrounding ions. The role lattice distortions plays on the $J$=0 state requires further investigation, in particular if it is a perturbation or acts to break the approximations of the $J$=0 state. Indeed this may offer a route to tune the $J$=0 state if the structure can be suitably altered through pressure, strain, chemical doping or other mechanisms. 

The concomitant behavior revealed here from the $A$ and $B$ sites offers several avenues of interest in Ho$_2$Os$_2$O$_7$. Unlike other spin ice candidates where the $B$ site is non-magnetic, the potential magnetism associated with the  Os$^{4+}$ ion adds a rare materials realization to investigate the interplay between these behavior on the $A$ and $B$ sites. Ho$_2$Ru$_2$O$_7$ offers some analogies, however, it was shown that both the Ru ion and then ultimately the Ho ion order, with strong evidence for magnetic interactions between the Ho and Ru sublattices. The exact nature of the magnetic ground state and excitations in  Ho$_2$Os$_2$O$_7$ is therefore an important open question for future theoretical and experimental studies. Indications of a phase transition at 36 K were associated with Os magnetic ordering. In a $J$=0 model this can occur via an excitonic route from exchange interactions, or alternatively from breaking the $J$=0 state due to trigonal distortions. However, concomitant short range ordering of the Ho ions occurs, indicating a strong interplay between the Os and Ho ions. This is analogous to the behavior in Ho$_2$Ru$_2$O$_7$. As shown here measurements that isolate each ion can provide unique information that will help build a complete understanding of Ho$_2$Os$_2$O$_7$ and related intriguing materials, with the interactions between these ions playing a significant role.

This research used resources at the High Flux Isotope Reactor, a DOE Office of Science User Facility operated by the Oak Ridge National Laboratory. This research used resources of the Advanced Photon Source, a U.S. Department of Energy (DOE) Office of Science User Facility operated for the DOE Office of Science by Argonne National Laboratory under Contract No. DE-AC02-06CH11357. JQY is supported by the US Department of Energy, Office of Science, Basic Energy Sciences, Materials Sciences and Engineering Division. This manuscript has been authored by UT-Battelle, LLC under Contract No. DE-AC05-00OR22725 with the U.S. Department of Energy. The United States Government retains and the publisher, by accepting the article for publication, acknowledges that the United States Government retains a non-exclusive, paidup, irrevocable, world-wide license to publish or reproduce the published form of this manuscript, or allow others to do so, for United States Government purposes. The Department of Energy will provide public access to these results of federally sponsored research in accordance with the DOE Public Access Plan(http://energy.gov/downloads/doepublic-access-plan).

\end{document}